\documentclass[letter]{aa} 
\usepackage{graphicx}
\usepackage{txfonts}
\usepackage{natbib}
\bibpunct{(}{)}{;}{a}{}{,}
\begin{document}
   \title{Analysis of the infrared spectra of the peculiar post-AGB stars EP\,Lyr and HD\,52961. 
\thanks{Based on observations made with the 1.2\,m Flemish Mercator telescope
at Roque de los Muchachos, Spain, the 1.2\,m Swiss Euler telescope at La Silla,
Chile and on observations made with the Spitzer Space Telescope (program id 3274), which is operated 
by the Jet Propulsion Laboratory, California Institute of Technology under a contract with NASA.
$^{**}$ NASA Postdoctoral Fellow}}

   \author{
          C. Gielen\inst{1}
          \and
          H. Van Winckel\inst{1}
          \and
          M. Matsuura \inst{2,3}
          \and
          M. Min \inst{4}
          \and
          P. Deroo$^{**}$ \inst{5}
          \and
          L.B.F.M. Waters \inst{1,4}
          \and
          C. Dominik \inst{4,6}
          }


   \institute{Instituut voor Sterrenkunde,
              Katholieke Universiteit Leuven, Celestijnenlaan 200D, 3001 Leuven, Belgium\\
	      \email{clio.gielen@ster.kuleuven.be}
              \and
              Astrophysics Group, Department of Physics and Astronomy, University College London, Gower Street, London WC1E 6BT, United Kingdom
              \and
              Division of Optical and IR Astronomy, National Astronomical Observatory of Japan, Osawa 2-21-1, Mitaka, Tokyo 181-8588, Japan
              \and
              Sterrenkundig Instituut 'Anton Pannekoek', 
              Universiteit Amsterdam, Kruislaan 403, 1098 Amsterdam, The Netherlands
              \and
              Jet Propulsion Laboratory, 4800 Oak Grove Drive, Pasadena, CA 91109, US 
              \and
              Department of Astrophysics, Radboud University Nijmegen, PO Box 9010, 6500 GL Nijmegen, The Netherlands
            }

   \date{Received ; accepted }

  \abstract
   {}
   {We aim to study in detail the peculiar mineralogy and structure of the circumstellar environment of two binary post-AGB stars,
EP\,Lyr and HD\,52961. Both stars were selected from a larger sample of evolved disc sources observed with Spitzer and show
unique solid-state and gas features in their infrared spectra. Moreover, they show a very small infrared excess in comparison with the other
sample stars.}
   {The different dust and gas species are identified on the basis of high-resolution Spitzer-IRS spectra. We fit the full spectrum to constrain 
    grain sizes and temperature distributions in the discs. This, combined with our broad-band spectral energy distribution and interferometric measurements, allows us to study the physical structure of
    the disc, using a self-consistent 2D radiative-transfer disc model.}
   {We find that both stars have strong emission features due to CO$_2$ gas, dominated by $^{12}$C$^{16}$O$_2$, but with clear $^{13}$C$^{16}$O$_2$ and
even $^{16}$O$^{12}$C$^{18}$O isotopic signatures. Crystalline silicates are apparent in both sources
but proved very hard to model. EP\,Lyr also shows evidence of mixed chemistry, with emission features of the rare class-C PAHs. 
Whether these PAHs reside in the oxygen-rich disc or in a carbon-rich outflow is still unclear. With the strongly processed silicates, the mixed chemistry and
the low $^{12}$C/$^{13}$C ratio, EP\,Lyr resembles some silicate J-type stars, although the depleted photosphere makes nucleosynthetic signatures difficult to probe.
We find that the disc environment of both sources is, to a first approximation, well modelled with a passive disc, but additional physics such as grain settling, radial dust distributions, and an outflow component must be included to explain the details of the observed spectral energy distributions in both stars. }
   {}
   \maketitle
%

\section{Introduction}
\label{introduction}

The infrared spectra of post-AGB stars are often characterised by
strong spectral signatures. These are formed in the gas and dust-rich
circumstellar environment (CE), which is a remnant of the strong mass
loss that occurred during the previous asymptotic giant branch (AGB)
evolutionary phase. The chemistry in this circumstellar environment is found to be
oxygen-rich or carbon-rich, depending on whether oxygen or carbon is
more abundant. The less abundant of the two will be locked in the
very stable CO molecule that forms in the stellar photosphere.

Typical post-AGB outflow sources that have O-rich CE not only show the
well-known 9.7 and 18\,$\mu$m features of amorphous silicates but also
narrower features, arising from crystalline silicates.
\citep[e.g.][]{waters96,molster02a}. The condensates in C-rich outflows show
features of carbon-species such as SiC, MgS or polycyclic aromatic
hydrocarbons (PAHs) \citep[e.g.][]{hony01,hony02,peeters02}. 
They are also characterised by an
often very strong feature at 21\,$\mu$m \citep{kwok89,volk99,hony03}.
The photospheres of these 21\,$\mu$m sources show strong enhancements of carbon and
s-process elements \citep[e.g.][]{reyniers04,reyniers07b} and the 21
$\mu$m stars are recognised as post-AGB carbon stars \citep[e.g.][]{vanwinckel00}.

Some evolved objects show, however, features of both O-rich and C-rich
dust species in their spectra. They are called mixed chemistry sources.
This chemistry is detected in several sources in a wide range of different
evolutionary stages. Some examples include Herbig Ae stars, or AGB stars, such as
J-type carbon stars with silicate dust emission
\citep{littlemarenin86,lloydevans90}. Others are red giants, for example
HD\,233517, an evolved O-rich red giant with PAHs in a circumstellar
disc \citep{jura06}. Other examples are planetary nebulae (PNe) with evidence of silicates and PAHs
\citep{kemper02,gutenkunst08}, or the hydrogen-poor [WC] central stars of PNe
\citep{waters98,cohen99}. Also some M supergiants are associated with emission due to PAHs
\citep{sylvester98,sloan08}.  

Post-AGB stars with evidence of mixed
chemistry include HD\,44179, the central star of the carbon-rich Red
Rectangle nebula \citep{cohen75}. The central star is a binary surrounded by a
Keplerian O-rich circumbinary disc \citep[e.g.][]{vanwinckel95,
waters98,menshchikov02,bujarrabal05}. Here the formation of the disc
is believed to have antedated the C-rich transition of the central
star \citep[e.g.][]{cohen04,witt08}.

Studies have shown that these evolved binaries with circumbinary discs are 
much more abundant than anticipated \citep{deruyter06,vanwinckel07}.
Interferometric studies
\citep{deroo06,deroo07b} prove that the discs are indeed very compact, with
radii around 50\,AU in the N-band. The discs are also the natural environment 
of the observed photospheric chemical depletion pattern in these stars
\citep{vanwinckel98,giridhar00}, due to chemical fractionation by dust
formation in the circumstellar environment \citep{waters92} and subsequent
accretion of the gas component. The presence of a long-lived
stable reservoir of dust grains also could allow for the observed strong processing of the
silicate dust grains, both in size as well as in crystallinity
\citep{molster02a,gielen08}.

Dusty RV\,Tauri stars are a distinct class in the post-AGB stars.
They cross the instability strip, and are therefore pulsating stars
\citep{jura86,jura99,deruyter05}. 
RV\,Tauri stars show large-amplitude photometric variations with alternating
deep and shallow minima. The members are located in the high-luminosity end of the population\,II instability strip, and the
photometric variations are interpreted as being due to radial
pulsations. Circumstellar dust emission was observed in many of them
\citep{jura86}, and this was generally acknowledged to be a decisive
character to place these stars in the post-AGB phase of evolution. 
The grains in almost all dusty RV\,Tauri stars are, however, not freely
expanding but likely also trapped in a disc \citep{vanwinckel99,deruyter05, deruyter06}.

In this paper we focus on two peculiar post-AGB stars with RV\,Tauri
pulsational characteristics: EP\,Lyr and HD\,52961.
These stars show unique spectral features and have very small
infrared excesses in comparison to the larger sample.

The outline of the paper is as follows: We start with a short
description of the programme stars in Sect.~\ref{progstars}. In
Sect.~\ref{observations} we give an overview of the different
observations and reduction strategies. The analysis based on the
Spitzer spectra is given in Sect.~\ref{spectralanalysis} and
subdivided in different subsections. Sect.~\ref{silicates} contains a
description of the silicate dust features and the modelling of the
Spitzer-IRS spectra.  The CO$_2$ gas features are discussed in
Sect.~\ref{co2} and the observed PAH features in EP\,Lyr in
Sect.~\ref{pah}. In Sect.~\ref{sed} we model the observed SEDs using
a passive disc model, also constrained with MIDI interferometric measurements. The discussion of our different results and our
conclusions are presented in Sect.~\ref{conclusions}.

\section{Programme stars}
\label{progstars}

In our previous study we described and modelled the Spitzer-IRS spectra of 
21 sources and found that the dust around these stars is all O-rich and on average highly crystalline \citep{gielen08}. 
The two stars discussed here have the lowest $L_{\rm IR}/L_*$, respectively 12\% and 3\%, in the larger Spitzer sample, where an average 
of about 50\% was found. 
The large infrared luminosity can be explained with a passive disc model, provided that the inner rim is close to the star
and the scale height of the disc is significant \citep[e.g.][]{deroo07b}. The low observed $L_{\rm IR}/L_*$ values of both stars point to a small disc scale height and/or a much larger inner gap, as it is unlikely that a disc is optically thin in the radial direction. 
Not only do they have the lowest $L_{\rm IR}/L_*$ values, both stars show
unique spectral signatures in comparison to the larger sample. We therefore selected these objects for a more detailed analysis.

\begin{table*}
\caption{List of stellar parameters for our sample sources.}
\label{sterren}
\vspace{0.5cm}
\hspace{0.cm}
\centering
\begin{tabular}{llrrcccrrclll}
\hline \hline
Name & $\alpha$ (J2000) & $\delta$ (J2000)  & $T_{\rm eff}$ & $\log g$ & [Fe/H] & P$_{\rm orb}$ & $e$ & $E(B-V)_{\rm tot}$ & $L_{\rm IR}/L_*$ & $d$\\ 
     & (h m s)          & ($^\circ$ ' '')   & (K)       & (cgs)    &        & (days)      &     &                & (\%)         & (kpc)\\
\hline
EP\,Lyr      & 19 18 17.5 & $+$27 50 38  & 7000 & 2.0 & -1.5   &       &      & $0.51\pm0.01$ & $3\pm0$ & $3.2\pm1.0$\\ 
HD\,52961    & 07 03 39.6 & $+$10 46 13  & 6000 & 0.5 & -4.8   & 1310  & 0.21 & $0.06\pm0.02$ & $12\pm1$ & $1.6\pm0.5$\\ 
\hline
\end{tabular}
\begin{footnotesize}
\begin{flushleft}
Note: Listed are the name, equatorial coordinates $\alpha$ and $\delta$(J2000), effective
temperature $T_{\rm eff}$, surface gravity $\log g$, and metallicity [Fe/H] of our sample stars.
For the model parameters we refer to \citet{deruyter06}. Also given are the orbital period and the
eccentricity \citep[see references in][]{deruyter06,gielen07}. The total reddening $E(B-V)_{\rm tot}$, the energy ratio $L_{\rm IR}/L_*$ and the calculated distance,
assuming a luminosity of $L_*=3000\pm2000$\,L$_{\odot}$.
\end{flushleft}
\end{footnotesize}
\end{table*}

\subsection{EP\,Lyr}

\citet{schneller31} discovered the variability of EP\,Lyr and classified it as an RVb star.
RVb stars are objects with a variable mean magnitude, in the General Catalogue of Variable
Stars \citep{kholopov99}. Other studies \citep{zsoldos95,gonzalez97a}
classify the light curve as an RVa photometric variable, having a constant mean
magnitude, with a period
of $P=83.46$ days. \citet{preston63} classify it as an RVB
spectroscopic variable. \citet{gonzalez97a} performed an abundance
analysis on EP\,Lyr where they deduced stellar parameters (see
Table~\ref{sterren}) and found the star to be metal-poor, oxygen-rich and severely
depleted. Using the molecular lines found in the spectra, they also
quantified the $^{12}$C/$^{13}$C ratio to be $9\pm1$. In the radial
velocity data there is also evidence that EP\,Lyr must have a stellar
mass companion, but additional observations are necessary to determine
the orbit.

\subsection{HD\,52961}

HD\,52961 is an RV\,Tauri like object, similar to class RVb objects
\citep{waelkens91b}, with a photometric variability of 72 days due to
clear radial pulsations \citep{waelkens91b}. The binarity of
HD\,52961 was first reported by \citet{vanwinckel95} and further
refined in \citet{vanwinckel99} and \citet{deroo06}, where an orbital period of
$P_{\rm orb}=1297\pm7$ days and an eccentricity of $e=0.22\pm0.05$ was
found. On top of the stable photometric variation due to the
pulsation, another long-term photometric variation was
detected, correlated with the orbital period. \citet{vanwinckel99} conclude that this can be understood as
caused by variable circumstellar extinction during the orbital motion.

The star is a highly metal-poor object with [Fe/H]\,$=$\,$-4.8$
\citep{waelkens91a} and has an extremely high zinc to iron ratio of
[Zn/Fe]\,$=$\,$+3.1$ \citep{vanwinckel92}.
The star is one of the most extremely depleted objects known.

HD\,52961 has been studied with mid-IR long-baseline interferometry
using the VLTI/MIDI instrument \citep{deroo06}. They find that
the dust emission originates from a very small but resolved region,
estimated to be $\sim35$\,mas at 8\,$\mu$m and $\sim55$\,mas at
13\,$\mu$m, likely trapped in a stable disc. The dust distribution
through the disc is not homogeneous: the crystallinity is
higher in the hotter inner region.

\section{Observations}
\label{observations}

High- and low-resolution spectra of 21 post-AGB stars were obtained
using the Infrared Spectrograph (IRS; \citealp{houck04}) aboard the
Spitzer Space Telescope \citep{werner04} in February 2005. The spectra
were observed using combinations of the short-low (SL), short-high
(SH) and long-high (LH) modules. SL ($\lambda$=$5.3-14.5$\,$\mu$m)
spectra have a resolving power of R=$\lambda/\bigtriangleup\lambda
\sim$ 100, SH ($\lambda$=$10.0-19.5$\,$\mu$m) and LH
($\lambda$=$19.3-37.0$\,$\mu$m) spectra have a resolving power of $\sim$
600. Exposure times were chosen to achieve an S/N ratio of around 400
for the high-resolution modes, which we complemented with short
exposures in low-resolution mode with an S/N ratio around 100, using
the first generation of the exposure time calculator of the call for
proposals.

The spectra were extracted from the SSC data pipeline version
S13.2.0 products, using the c2d Interactive Analysis reduction
software package \citep{kessler06,lahuis06}. This data processing
includes bad-pixel correction, extraction, defringing and order
matching. To match the different orders, we applied small scaling
corrections.

\begin{figure}
\vspace{0cm}
\hspace{0cm}
\centering
\resizebox{9cm}{!}{\includegraphics{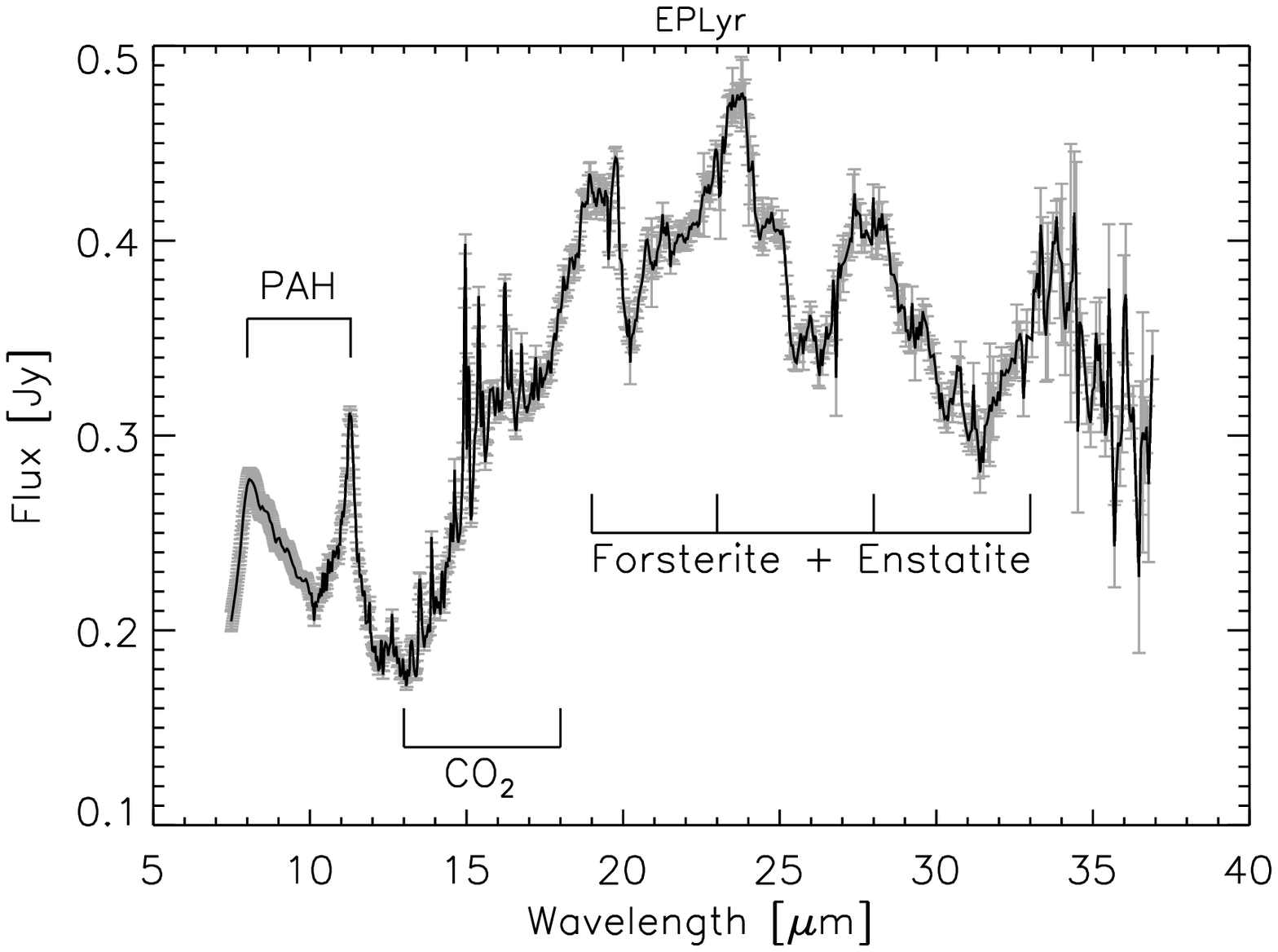}}
\vspace{0.5cm}
\hspace{0cm}
\resizebox{9cm}{!}{\includegraphics{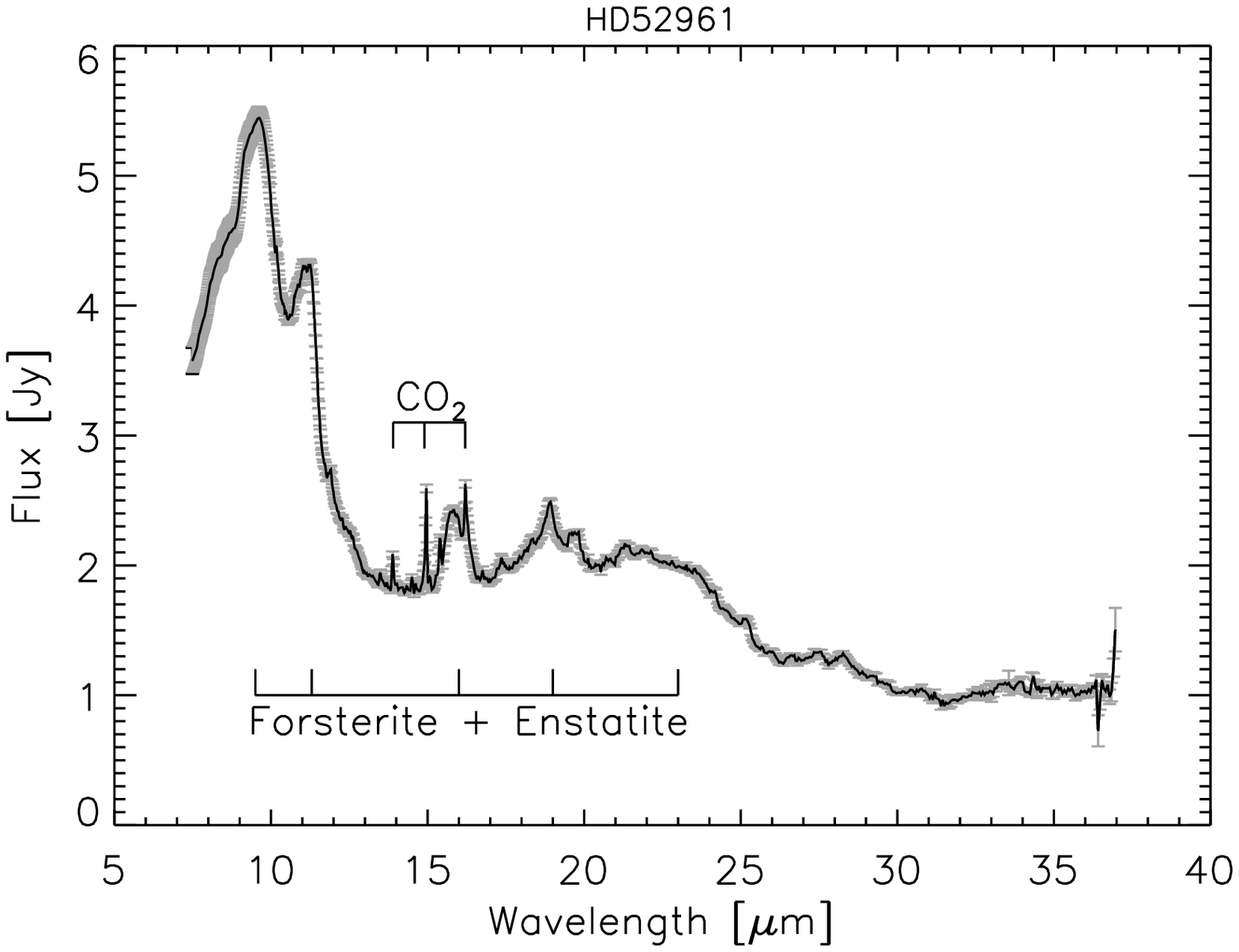}}
\caption{The combined Spitzer-IRS high- and low-resolution spectrum of EP\,Lyr (top) and HD\,52961 (bottom).
In grey we overplot the estimated errors.}
\label{eplyr_hd52961}%
\end{figure}

\section{Spectral analysis}
\label{spectralanalysis}

\subsection{General}

A look at the spectra of EP\,Lyr and HD\,52961
(Fig.~\ref{eplyr_hd52961}) show very rich spectra with 
quite different continuum slopes. EP\,Lyr shows strong emission features at
longer wavelengths, with peak emission in the 20\,$\mu$m region,
whereas HD\,52961 is characterised by a strong 10\,$\mu$m emission
feature on top of a much steeper continuum.

Common dust species found in oxygen-rich post-AGB stars are amorphous
silicates, namely olivine and pyroxene.
Amorphous olivine (Mg$_{2x}$Fe$_{2(1-x)}$SiO$_4$, where $0 \leq x \leq
1$ denotes the magnesium content) has very prominent broad features
around 9.8\,$\mu$m and 18\,$\mu$m. Amorphous pyroxene
(Mg$_{x}$Fe$_{1-x}$SiO$_3$) shows a 10\,$\mu$m feature similar to that
of amorphous olivine, but shifted towards shorter wavelengths. Also
the shape of the 18\,$\mu$m feature is slightly different. 
For EP\,Lyr it is unclear whether there is a significant contribution of
amorphous silicates. Small amorphous silicates could contribute to the observed strong emission bump at
20\,$\mu$m in EP\,Lyr, but as there does not seem to be a 10\,$\mu$m amorphous feature, the 20\,$\mu$m
bump could be purely continuum dominated. HD\,52961
has clear strong emission of amorphous silicates at 10\,$\mu$m, but the 
profile shows complex narrow subfeatures. Very little contribution at 20\,$\mu$m is seen.

Both stars show strong narrow emission features which can be
identified as being due to crystalline silicates. The Mg-rich end
members of crystalline olivine and pyroxene, forsterite
(Mg$_2$SiO$_4$) and enstatite (MgSiO$_3$), show strong but narrow
features at distinct wavelengths around $11.3 - 16.2 - 19.7 - 23.7 -
28$ and 33.6\,$\mu$m. For EP\,Lyr the silicate emission only clearly
starts longward of 18\,$\mu$m, where strong emission features around
$19-23-27$ and 33\,$\mu$m can be seen. HD\,52961 has strong narrow
features at $9.8-11.3$\,$\mu$m and a remarkably strong 16\,$\mu$m feature.  
If this strong 16\,$\mu$m band is only due to forsterite 
it has shifted considerably to shorter
wavelengths. A significant 16\,$\mu$m feature is seen in several evolved
disc sources but it is never as strong as in HD\,52961 \citep{gielen08}.  

EP\,Lyr shows evidence of the presence of carbon-rich dust species
with probable PAH identifications at 8.1 and 11.3\,$\mu$m. The detection of
PAH emission together with silicates is surprising and only observed
in a few other post-AGB sources. The analysis of the PAH
features is given in Sect.~\ref{pah}.

\begin{figure}
\centering
\resizebox{8cm}{!}{\includegraphics{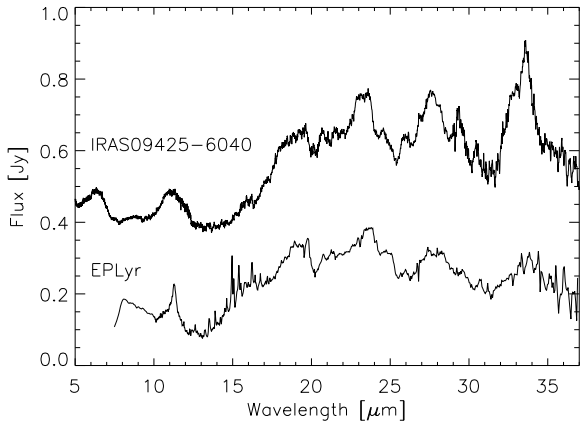}}
\caption{The Spitzer-IRS spectrum of EP\,Lyr compared to the ISO-SWS spectrum of IRAS\,09425-6040.
The spectrum of IRAS\,09425-6040 is normalised and offset for comparison.}
\label{eplyr-iras09425}
\end{figure}

The spectrum of EP\,Lyr shows a strong resemblance to that of IRAS\,09245-6040 (Fig.~\ref{eplyr-iras09425}),
a silicate J-type carbon AGB star \citep{molster01,garciahernandez06}.
Silicate J-type carbon stars have surprisingly low $^{12}$C/$^{13}$C ratios and do not show the typical s-process
overabundances seen in N-type carbon stars \citep{abia00}. The infrared spectrum of these stars shows features of both carbon- and oxygen-rich 
dust species. 

Of the silicate J-type carbon stars, only 10\% show emission bands due to crystalline material \citep{lloydevans91,ohnaka99}.
The formation history of these stars is still unclear, but the most promising
scenario for the presence of silicates in these stars, is that they are binaries with an undetected companion \citep{lloydevans90,yamamura00}.
A disc is supposed to be formed when the primary was still an O-rich giant. After that the star
underwent thermal pulses and evolved into a carbon star. 
The silicate disc could be either captured from the wind \citep{mastrodemos99} or the result
of a phase of strong binary interaction in a narrow system.
Systems with strong crystalline features in their spectra, such as IRAS\,09425-6040 or IRAS\,18006-3213 \citep{deroo07a},
would then be a result of mass-transfer into a circumbinary system, whereas sources dominated by amorphous silicates, such as
V778\,Cygni \citep{yamamura00} or BH\,Gem \citep{ohnaka08},
consist of a wide binary with a circumcompanion disc.
To date, no orbits are known, however, and direct evidence of binarity is found in a few objects only \citep{izumiura08}.

For IRAS\,09245-6040, the $^{12}$C/$^{13}$C ratio is calculated to be $15\pm6$ \citep{garciahernandez06}.
In the ISO-SWS spectrum features of C$_2$H$_2$, HCN, CO, C$_3$ and SiC are seen shortward of 15\,$\mu$m;
after 15\,$\mu$m the spectrum is dominated by strong emission features of Mg-rich crystalline silicates \citep{molster01}.
As in EP\,Lyr, there is no evidence of a strong contribution of amorphous silicates.

Finally in both EP\,Lyr and HD\,52961, clear CO$_2$ gas emission features are
detected the $13-18$\,$\mu$m region. This is discussed in Sect.~\ref{co2}.

\begin{figure}
\vspace{0cm}
\hspace{0cm}
\centering
\resizebox{9cm}{!}{ \includegraphics{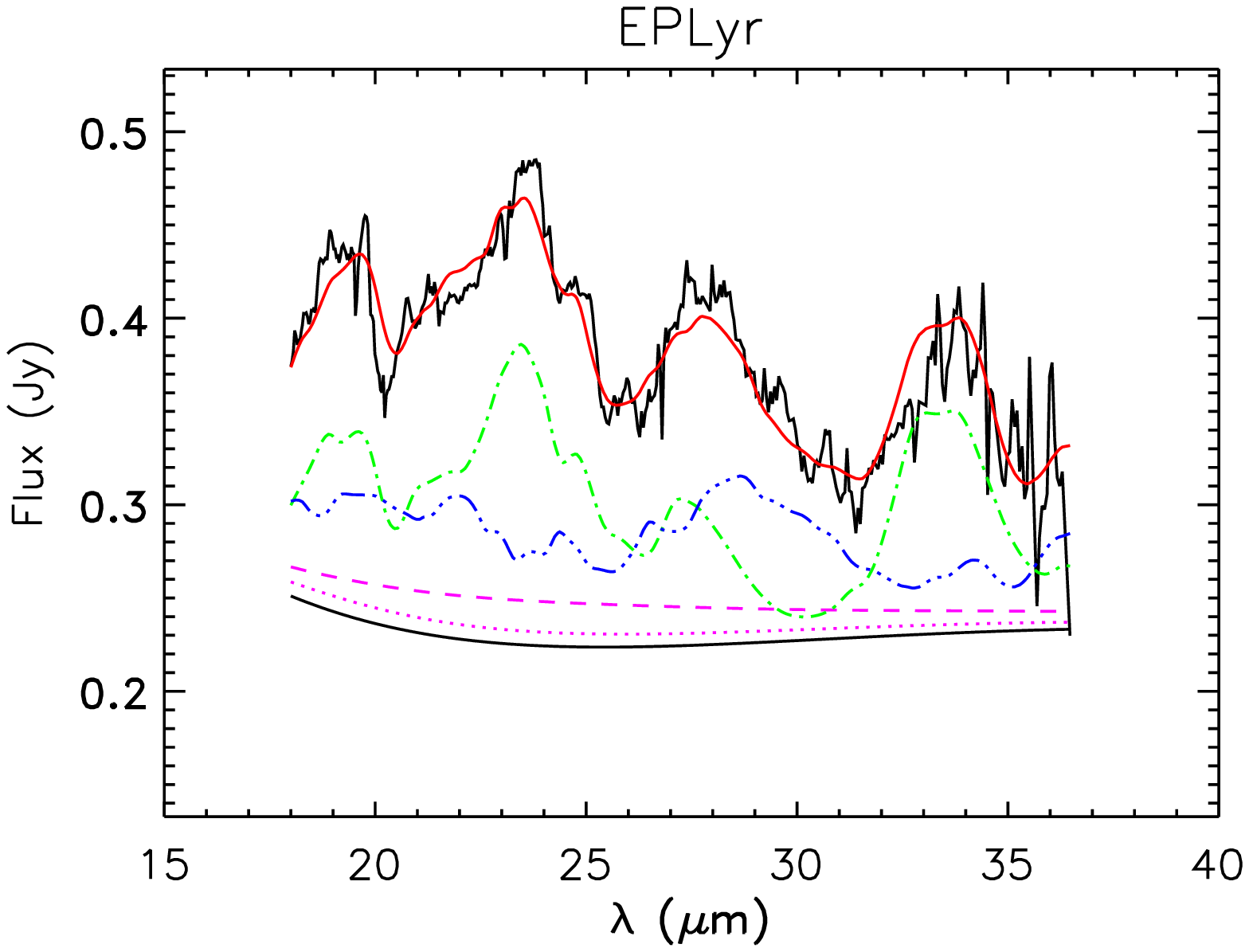}}
\resizebox{9cm}{!}{ \includegraphics{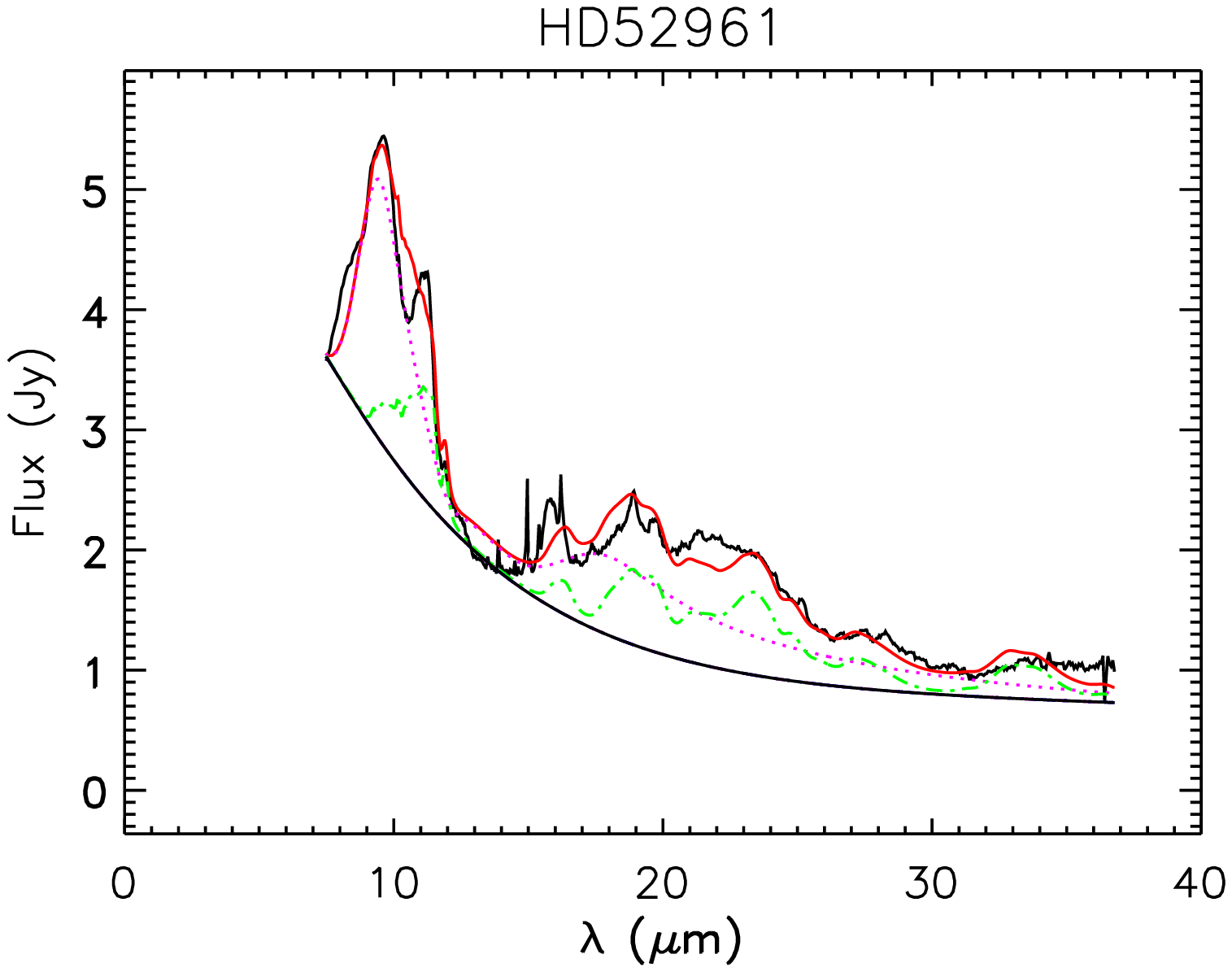}}
\caption{Best fits for EP\,Lyr and HD\,52961.
The observed spectrum (black curve) is plotted together with the best model fit (red curve) and the continuum (black solid line).
Forsterite is plotted in dash-dot lines (green) and enstatite in dash-dot-dotted lines (blue).
Small amorphous grains are plotted as dotted lines (magenta) and large amorphous grains as dashed lines (magenta).}
\label{silicate_fit}%
\end{figure}

\subsection{Silicate dust emission}
\label{silicates}

We optimised the fitting procedure as discussed in \citet{gielen08} for these two outliers,
where we modelled the full Spitzer sample, consisting of 21 stars. 
In short we assume the flux to be originating from an optically thin region,
so we can make linear combinations of the absorption profiles to calculate
the model spectrum.
In our previous modelling we found that, on average for the full sample, the best fit was obtained
using relatively large grains ($\geq 2$\,$\mu$m) in an irregular
Gaussian Random Fields (GRF) dust model. For EP\,Lyr and
HD\,52961 however, we already found that using smaller grain sizes
($\leq 2$\,$\mu$m) improved the fit considerably.

So we repeated the analysis for EP\,Lyr and HD\,52961, allowing for
different dust shapes, grain sizes and Mg/Fe content in the amorphous
grains. We tested Mie, GRF and DHS (Distribution of Hollow Spheres)
dust models in grain sizes ranging from 0.1 to
4.0\,$\mu$m. In order to test for the presence of Fe-poor amorphous
dust, we perform the modelling both with pure Mg-rich amorphous
silicates ($x=1$) and with the more standard Mg-Fe amorphous silicate
dust ($x=0.5$). For a detailed description of the fitting routine we
refer to \citet{gielen08}. The results of the fitting can be found in
Table~\ref{chi2}. As for EP\,Lyr the silicate signatures only
appear after 18\,$\mu$m; we only fit this part of the Spitzer
spectrum.

\begin{table}
\caption{$\chi^2$ values for different models used in our full spectral fitting.}
\label{chi2}
\centering
\vspace{0.5cm}
\hspace{0.cm}
\begin{tabular}{llll}
\hline
\hline
       & EP\,Lyr & HD\,52961 & model description \\   
       &  $  \chi^2$  &  $\chi^2$  & \\
\hline
model1 &  21.7   & 129.4     & Mie - $0.1-2.0$\,$\mu$m - $x=0.5$    \\
model2 &  6.2  &  67.5  & DHS - $0.1-1.5$\,$\mu$m - $x=0.5$    \\  
model3 &  6.2  &  63.8  & DHS - $0.1-1.5$\,$\mu$m - $x=1.0$    \\   
model4 &  8.4  &  101.4  & DHS - $1.5-3.0$\,$\mu$m - $x=0.5$    \\  
model5 &  8.6  &  140.8  & DHS - $1.5-3.0$\,$\mu$m - $x=1.0$    \\  
model6 &  5.9   & 64.2      & GRF - $0.1-2.0$\,$\mu$m - $x=0.5$    \\
model7 &  6.3   & 50.0      & GRF - $0.1-2.0$\,$\mu$m - $x=1.0$    \\
model8 &  5.8   & 96.5      & GRF - $2.0-4.0$\,$\mu$m - $x=0.5$    \\
model9 &  5.4   & 72.2      & GRF - $2.0-4.0$\,$\mu$m - $x=1.0$    \\
\hline
\end{tabular}
\begin{footnotesize}
\begin{flushleft}
Note: For each model we give the used dust approximation, grain size and Mg-Fe content in the amorphous grains.
$x=1.0$ denotes pure Mg-rich amorphous dust, $x=0.5$ the more standard Mg-Fe amorphous silicates.
\end{flushleft}
\end{footnotesize}
\end{table}

The $\chi^2$ values of our fitting (Table~\ref{chi2}) are still quite
high for HD\,52961 but, confirming the result of \citet{gielen08}, 
we can already tell that for both stars Mie theory is not a
good dust approximation. For EP\,Lyr the GRF grains prove the best
match, but the difference in $\chi^2$ with the small DHS grain
approximation is only minimal. The best fit to EP\,Lyr is obtained using
both small (0.1\,$\mu$m) and larger (2.0\,$\mu$m) silicate grains.
The small difference in calculated $\chi^2$ values for EP\,Lyr is due
to the low signal-to-noise ratio, making it hard to distinguish between different synthetic
emission profiles.
For HD\,52961 small grains in Mg-rich silicates give the best
$\chi^2$. Plots of our best fitting models can be found in
Fig.~\ref{silicate_fit}. Table~\ref{fitresults1} gives the resulting parameters.

The large $\chi^2$ value of HD\,52961 quantifies that this star has a very peculiar, 
unique chemistry, and we did not succeed
in explaining all of the observed features.
The strong forsterite 11.3\,$\mu$m feature in the
GRF dust approximation is clearly too broad. DHS grains fit the feature better, but other feature profiles are fitted less well
with this approximation. There also appears to be a short wavelength shoulder on
the amorphous 9.8\,$\mu$m feature, which is not explained in the
modelling. The strong 16.5\,$\mu$m feature is not reproduced in
central wavelength by any of the different models. We already
observed this trend in our full sample fitting \citep{gielen08}, where
the feature seemed to be shifted bluewards in comparison with the mean
spectrum of the full sample. The two narrow features around
19\,$\mu$m could be an artifact of the data reduction, since in this
region there can be a bad overlap between the SH and LH
Spitzer-IRS high-resolution bands.

For EP\,Lyr we fit the spectrum longwards of 18\,$\mu$m, where the silicate features are seen. This gives
dust temperatures between 100 and 230\,K.
This model, however, does not fit the spectrum before 18\,$\mu$m, since the continuum does not follow the observed strong downward slope before 20\,$\mu$m.
If we try to fit the full Spitzer wavelength range we find we can get a better fit to the underlying continuum but then
the features at 27 and 33\,$\mu$m are much stronger in the observed spectrum than in our best model. Unlike in other sources, a two temperature approach fails to model both the observed continuum and the coolest features for the full Spitzer wavelength spectrum of EP\,Lyr. Irrespective of the derived continuum temperature, all the tested models give estimates of the dust temperatures between $100-300$\,K, which agrees with the temperatures derived in the SED modelling (Sect.~\ref{sed}). Clearly, the crystalline dust particles must be quite cold.

\begin{table*}
\caption{Listed are the best fit parameters deduced from our full spectral fitting. }
\label{fitresults1}
\centering
\vspace{0.5cm}
\hspace{0.cm}
\begin{tabular}{lrllllll}
\hline \hline
Name & $\chi^2$ &  $T_{dust1}$ & $T_{dust2}$ & Fraction & $T_{cont1}$ & $T_{cont2}$ & Fraction \\
    &          &     (K)     & (K)         & $T_{dust1}$- $T_{dust2}$    & (K)         & (K)         & $T_{cont1}$-$T_{cont2}$   \\
\hline
EP\,Lyr & 5.4 &$ 114_{  14}^{  122}$ &$ 228_{  89}^{ 488}$ &$ 0.9_{ 0.6}^{ 0.1}- 0.1_{ 0.1}^{ 0.6}$ &$ 205_{ 103}^{  702}$ &$ 641_{ 301}^{  331}$ &$ 0.96_{ 0.06}^{ 0.02}- 0.04_{ 0.02}^{ 0.06}$\\
HD\,52961  & 50.0 &$ 200_{  0}^{  10}$ &$ 724_{  96}^{ 186}$ &$ 0.9_{ 0.1}^{ 0.0}- 0.1_{ 0.0}^{ 0.1}$ &$ 111_{ 11}^{  356}$ &$ 996_{ 111}^{  4}$ &$ 0.99_{ 0.03}^{ 0.00}- 0.01_{ 0.00}^{ 0.03}$\\
\end{tabular}

\vspace{0.5cm}
\hspace{0.cm}
\begin{tabular}{lccccc}
\hline \hline   
Name & Olivine & Pyroxene & Forsterite & Enstatite & Continuum\\
      & Small  -  Large & Small  -  Large &  Small  -   Large & Small  -   Large &\\
\hline
EP\,Lyr    &$ 6_{ 6}^{ 41}    -   5_{ 5}^{25}$    &$8_{7}^{34}    -  6_{6}^{23}$    &$34_{15}^{19}    -   9_{ 8}^{33}$    &$ 5_{ 5}^{20}    -  28_{21}^{17}$    &$53_{ 20}^{11}$\\

HD\,52961    &$ 0_{ 0}^{ 13}    -   1_{ 1}^{41}$    &$55_{18}^{12}    -  2_{2}^{43}$    &$6_{5}^{18}    -   33_{ 18}^{13}$    &$ 1_{ 1}^{8}    -  3_{3}^{26}$    &$69_{ 4}^{ 4}$\\

\hline
\end{tabular}
\begin{footnotesize}
\begin{flushleft}
Note: Top part: The $\chi^2$, dust and continuum temperatures and their
relative fractions. Bottom part: The abundances of small (0.1\,$\mu$m) and large (2.0\,$\mu$m) grains of the various
dust species are given as fractions of the total mass, excluding the dust responsible for the continuum emission.
The last column gives the continuum flux contribution, listed as a percentage of the total integrated flux over the 
full wavelength range. The errors were obtained using a Monte-Carlo
simulation based on 100 equivalent spectra. Details of the modelling
method are explained in \citet{gielen08}.
\end{flushleft}
\end{footnotesize}

\end{table*}  

\subsection{CO$_2$ emission}
\label{co2}

\subsubsection{Introduction}

CO$_2$ emission has been found in approximately 30\% of all O-rich AGB stars \citep{justtanont98, ryde99, sloan03}, 
but CO$_2$ detections in post-AGB stars
are rare. To our knowledge CO$_2$ gas has been found in only two post-AGB stars, the Red Rectangle
and HR\,4049 \citep{waters98, cami01}, which are also binaries
surrounded by a stable circumstellar disc. HR\,4049 is the only example of a post-AGB star 
showing CO$_2$ in emission in the $13-16$\,$\mu$m region.
\citet{cami01} argued that the isotopic distribution of oxygen in HR\,4049 is abnormal,
based on the isotope ratio analysis of the CO$_2$ emission features. This was not confirmed
by \citet{hinkle07}, who use high-resolution spectra of the fundamental and first overtone CO
vibro-rotational transition in the near-IR.

\begin{figure}
\centering
\resizebox{8cm}{!}{\includegraphics{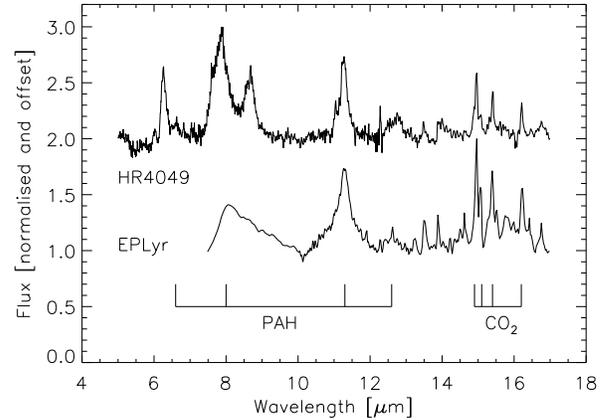}}
\caption{Comparison between the continuum-subtracted spectrum of EP\,Lyr and HR\,4049, another mixed chemistry source.
Both stars show clear emission due to CO$_2$ gas around 15\,$\mu$m and strong PAH features before 13\,$\mu$m.
The CO$_2$ features at $14.9-16.2$\,$\mu$m, 15.3\,$\mu$m and 15.1\,$\mu$m are respectively due to $^{12}$C$^{16}$O$_2$, $^{13}$C$^{16}$O$_2$
and $^{16}$O$^{12}$C$^{18}$O.
The PAH emission is clearly very different in HR\,4049, where class B PAHs are found, whereas in EP\,Lyr the
PAH features can be attributed to class C (see Sect.~\ref{pah}). }
\label{eplyr_hr4049}
\end{figure}

Both EP\,Lyr and HD\,52961 show clear gas phase emission lines of $^{12}$CO$_2$ and $^{13}$CO$_2$. 
These emission lines were also seen in only one other
source in our Spitzer sample \citep{gielen08}, namely in IRAS\,10174-5704. 

The CO$_2$ emission of EP\,Lyr seems to be lying on top of a ``plateau'' that extends from 13 to 17\,$\mu$m.
A similar plateau in this region is observed in PAH-rich sources \citep{peeters04}, but this plateau is much broader and ranges from
15 to 20\,$\mu$m and is often characterised by strong emission features at 16.4\,$\mu$m (and less prominent at 15.8, 17.4 and
19\,$\mu$m), and thus quite different to the one seen in EP\,Lyr. 

\subsubsection{Analysis}

\begin{figure}
\centering
\resizebox{8cm}{!}{\includegraphics{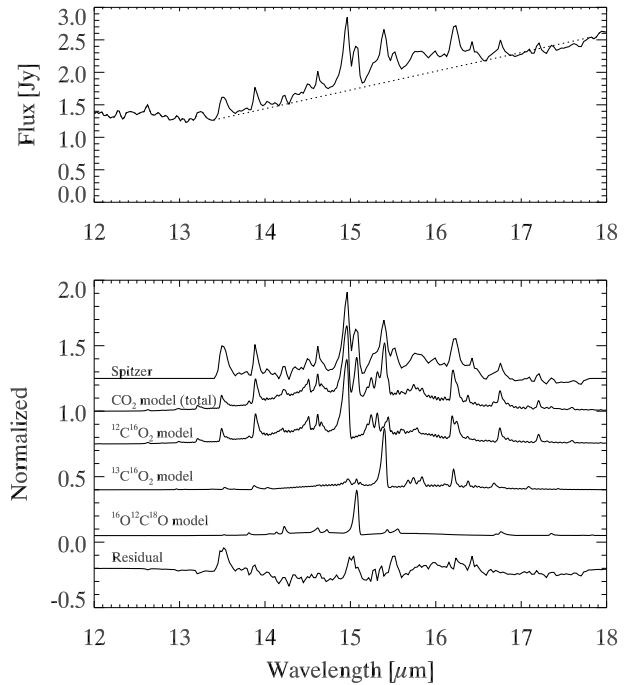}}
\caption{ EP\,Lyr: top panel shows the observed spectrum (solid line) and
pseudo-continuum spectra (dotted line).
The bottom panel shows the continuum divided spectrum, CO$_2$ model
spectra (combining all of the isotopes) and individual CO$_2$ isotope spectra
(from top to bottom $^{12}$C$^{16}$O$_2$, $^{13}$C$^{16}$O$_2$, $^{16}$O$^{12}$C$^{18}$O).
}
\label{Fig-EPLyr}
\end{figure}

\begin{figure}
\centering
\resizebox{8cm}{!}{\includegraphics{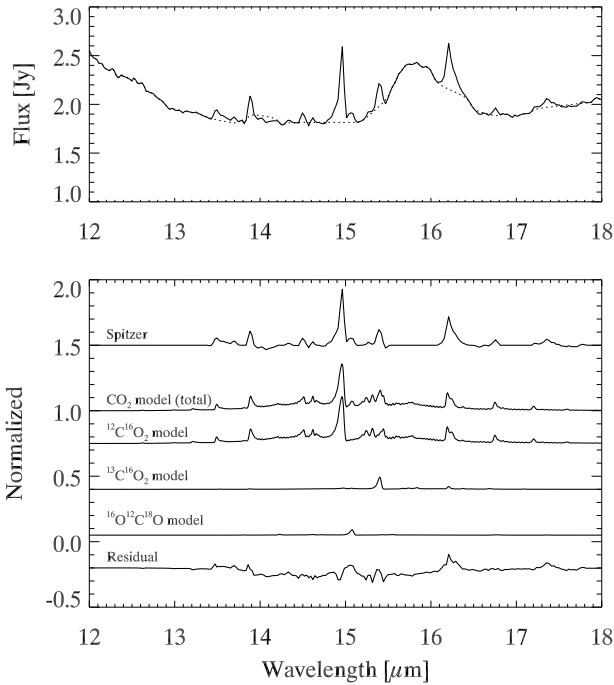}}
\caption{Same as Fig.~\ref{Fig-EPLyr} but for HD\,52961.
}
\label{Fig-HD52961}
\end{figure}

To retrieve the very rich spectral information of the CO$_2$ emission bands, we calculate spectra of CO$_2$, using HITRAN
line lists \citep{rothman05} and a circular slab model for the radiative-transfer \citep{matsuura02}.
The model has four parameters:
the excitation temperature ($T_{\rm ex}$), 
the total CO$_2$ column density ($N$),
the radius of CO$_2$ gas ($r$) and the isotope ratio.
The radius of the CO$_2$ layer is given relative to the radius of the background continuum source at 13\,$\mu$m.
The dependence of the CO$_2$ model spectra on these parameters are described by \citet{cami01}.
We estimate a pseudo-continuum by using a spline fit and a linear fit for HD\,52961
and EP\,Lyr, respectively. For HD\,52961, we interpolate the spectrum at the spectral range where CO$_2$
bands have little influence on the observed spectrum. The continuum was also chosen so that the forsterite
feature at 16\,$\mu$m would be removed.
A spline fit was tested for EP\,Lyr but failed because of 
the richness of CO$_2$ features in the mid-infrared range, so we simply use a linear interpolation between 13.4 and 17.8\,$\mu$m.
Estimated continua are displayed as dotted lines in the top panels of Figs.~\ref{Fig-EPLyr}
and \ref{Fig-HD52961}.
The resulted parameters for the model calculations are summarised in Table~\ref{parameters},
and the resulted spectra show the identifications of the different CO$_2$ bands
(bottom panels of Figs.~\ref{Fig-EPLyr} and \ref{Fig-HD52961}).

Many small features in EP\,Lyr in the $13.5-17$\,$\mu$m region are due to CO$_2$:
features at 13.5, 13.9, 14.7, 14.9, 16.2\,$\mu$m are attributed to the CO$_2$ main isotope $^{12}$C$^{16}$O$_2$.
The main isotopic $^{12}$C$^{16}$O$_2$ bands are probably optically thick, surpressing
the line intensities. $^{13}$C$^{16}$O$_2$ and
$^{16}$O$^{12}$C$^{18}$O bands are found at 15.3\,$\mu$m and 15.1\,$\mu$m, respectively.
The prominent $^{16}$O$^{12}$C$^{18}$O feature is surprising. This feature was also found
in the other binary post-AGB star HR\,4049 \citep{cami01}.

The model uses a high fraction of isotopes, but actual abundance ratios remain largely uncertain,
mainly because of the uncertainty of the interpolated continuum spectrum and the optical thickness
of the main isotope.
Nevertheless, these two features are particularly prominent in the spectrum of EP\,Lyr, more than in HD\,52961,
suggesting different isotope ratios for EP\,Lyr. We see that the observed ``plateau'' in EP\,Lyr can be explained by
the richness of the $^{12}$C$^{16}$O$_2$ features, but the $^{12}$C/ $^{13}$C ratio is confirmed to be low.
The strength of the isotopes, including the very rare $^{18}$O (in the Sun $^{16}$O/$^{18}$O is $\sim$\,500), is an exclusive feature of post-AGB stars.

%

\begin{table}
 \caption{Resulting parameters for the model calculations. The isotope ratio 
is as follows: $^{12}$C$^{16}$O$_2$:$^{13}$C$^{16}$O$_2$:$^{16}$O$^{12}$C$^{18}$O.  
\label{parameters}}
\centering
\vspace{0.5cm}
\hspace{0.cm}
\begin{tabular}{lrrrlrrrrccccccccc} \hline\hline
  & $T_{\rm ex}$ & $N$ & $r$ & isotope ratio \\
  &  (K)         & (cm$^{-2}$)    &  ($R_*$)   & \\ \hline
EP\,Lyr        & 900 & $8\times10^{18}$ & 4.7 & 0.7 : 0.2 : 0.1 \\
HD\,52961 & 800  & $5\times10^{18}$ & 4.7 & 0.93 : 0.05 : 0.02 \\
\hline
\end{tabular}
\end{table}

\subsection{PAH features}
\label{pah}

Polycyclic aromatic hydrocarbons are found in a large variety of
objects, including the diffuse ISM, HII regions, young stellar
objects, post-AGB stars and planetary nebulae. They have strong emission
features in the 3-13\,$\mu$m region \citep[e.g.][]{tielens08}. The feature at 3.3\,$\mu$m is
arises from the C-H stretching mode of neutral PAHs. The C-C modes
produce features with typical central wavelengths at 6.2 and
7.7\,$\mu$m. The 8.6\,$\mu$m feature is due to C-H in-plane bending
modes and features longward of 10\,$\mu$m can be attributed to C-H
out-of-plane bending modes.

\citet{peeters02} defined three groups of PAH spectra based on their
emission profiles and peak positions. The ``class A'' sources have
features at 6.22, 7.6 and 8.6\,$\mu$m. ``Class B'' sources show the
same features but shifted to the red, peaking at 6.27, 7.8 and
$>8.6$\,$\mu$m. They also identified two ``class C'' sources,
the Egg Nebula (AFGL\,2688) and IRAS\,13416-6243, both post-AGB
objects. These rare ``class C'' sources  show emission features at
6.3\,$\mu$m, no emission near 7.6\,$\mu$m, and a broad feature
centred around 8.2\,$\mu$m, extending beyond 9\,$\mu$m.

\begin{figure}
\vspace{0cm}
\hspace{0cm}
\centering
\resizebox{9cm}{!}{ \includegraphics{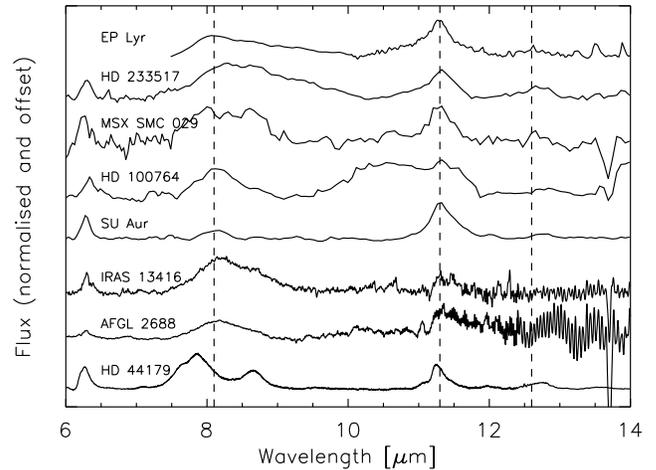}}
\caption{Continuum-subtracted spectra (based on a spline fit) of class-C sources as described in \citet{sloan07}.
The vertical dashed lines are at 8.1, 11.3 and 12.6\,$\mu$m. For comparison we also plot HD\,44179,
which has class B PAH emission.}
\label{splinesub}%
\end{figure}

With the release of the IRS aboard the Spitzer Space Telescope, a
limited number of additional class-C sources were discovered (Fig.~\ref{splinesub}). 
MSX SMC 029, a class-C post-AGB star in the SMC, was detected by
\citet{kraemer06}. \citet{sloan07} report on the detection of class-C
PAH features in HD\,100764, a carbon-rich red giant with evidence of
a circumstellar disc. \citet{jura06} also report on the detection of
class-C PAH features in a circumstellar disc around the oxygen-rich
K-giant HD\,233517. Two young objects, the T\,Tauri star SU\,Aur
\citep{Furlan06} and the Herbig Ae/Be source HD\,135344
\citep{sloan05}, also show PAH spectra of class C, although in HD\,135344
the PAH features seem to be somewhat more in between B and C. 
This source is also slightly hotter than other class C sources.
A comparison of the PAH features in all these sources is given by
\citet{sloan07}. They find that all the known class-C spectra are
excited by relatively cool stars of spectral type F or later and argue
that the hydrocarbons in these sources have not been exposed to much
ultraviolet radiation. The class-C PAHs are then relatively protected
and unprocessed, while class A and B PAHs have been exposed to more
energetic photons and are hence more processed.

\begin{figure}
\vspace{0cm}
\hspace{0cm}
\centering
\resizebox{9cm}{!}{ \includegraphics{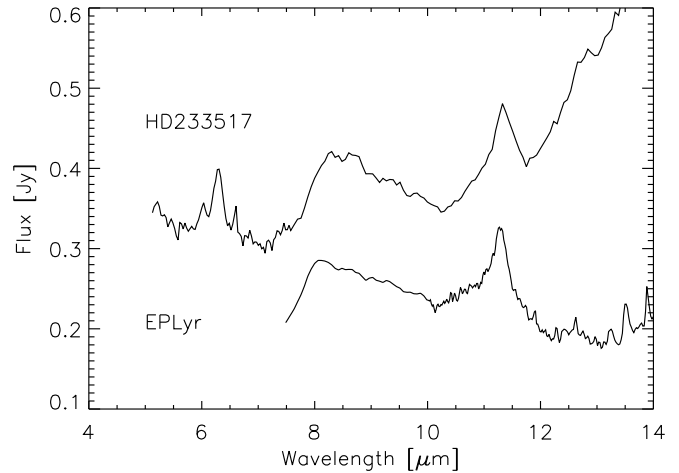}}
\caption{Comparison between the Spitzer-IRS spectrum of HD\,233517 and EP\,Lyr in the
6-14\,$\mu$m region.}
\label{hd233517}%
\end{figure}

EP\,Lyr shows PAH emission bands at 8.1, 11.3 and 12.6\,$\mu$m. Other
features can be seen at 13.25\,$\mu$m. Whether these can be
attributed to PAHs remains uncertain. The PAH spectrum of 
EP\,Lyr can be classified as class C.
EP\,Lyr is a high amplitude-variable with an effective temperature around
7000\,K, which is on the hot end of the other class-C emitters \citep[see][]{sloan07}.

Sofar, it is not clear whether the PAH carriers reside in the circumbinary disc, 
or in bipolar lobes created by
a more recent mass-loss event, as observed in HR\,4049
\citep{johnson99,dominik03,antoniucci05,hinkle07,menut09} and in the Red
Rectangle \citep{menshchikov02,cohen04}. For HR\,4049,
\citet{johnson99} found that the optical polarisation seems to
vary with orbital phase, suggesting that the polarisation in the
optical is due to scattering in the circumbinary disc. In the UV, the
polarisation is caused by scattering in the bipolar lobes, which
should contain a population of small grains, including the PAH
carriers. HR\,4049 and HD\,44179, as well as EP\,Lyr and HD\,52961, are
strongly depleted, and molecules or the formation of dust in these very
depleted environments is likely very different from solar
condenstation. As the CO molecule is abundant in the circumstellar
environment, accretion of circumstellar gas will likely result in a C$\sim$O photosphere.
HR\,4049 and HD\,44179 are stars with PAH emission belonging to the more standard
class B. 

Unlike what is detected around HD\,44179 (the ERE nebula) and HR 4049 (nano-diamonds), 
for EP\,Lyr the PAHs are the only carbon-rich component observed in the circumstellar spectra. 
There is no evidence that the photosphere of EP\,Lyr is in, or went through, a carbon-rich phase. 
The photosphere is depleted so that nucleosynthetic yields are very hard to recognise, 
but the photosphere is clearly oxygen-rich \citep{gonzalez97a}. 
A scenario involving hot-bottom burning to return to an oxygen-rich condition after a carbon-rich phase 
in the stellar evolution along the AGB is very unlikely:  it would imply the object is of a more massive origin \citep{lattanzio96,mcsaveney07}, 
which is in contradiction with its Galactic coordinates of $b = 6.9^{\circ}$. Assuming 10\,000\,L$_\odot$ for a putative massive progenitor, 
the distance above the Galactic plane would be about 660\,pc, which is very high for an object of $5-6$\,M$_\odot$. 
Moreover, the initial metallicity is likely subsolar as indicated by the solfur and
zinc abundances. Although, with the depletion, it is unclear whether these 
abunances of the volatiles do indeed represent the initial conditions. 
Also the pulsation period of EP\,Lyr is similar to other RV\,Tauri objects which are of low initial mass. 
We conclude that the photosphere of EP\,Lyr is now O-rich, and we argue it is very unlikely that it 
was ever in its history in a carbon-rich phase. The PAH synthesis likely occurred in O-rich conditions. 

As Figure~\ref{hd233517} shows,
the spectrum of EP\,Lyr has a striking resemblance to that of
HD\,233517, shortward of 13\,$\mu$m \citep{jura06}. 
\citet{jura03} hypothesises that HD\,233517 was a short-period binary on the main
sequence. A circumstellar disc was then formed when the companion star
was engulfed by the more massive star when it entered giant evolution,
followed by a phase of strong mass ejection in the equatorial plane.  
Since HD\,233517 is an oxygen-rich star, it
is remarkable that the disc shows features of carbon-rich chemistry.
\citet{jura06} propose a scenario in which the PAHs could be formed
inside the disc due to Fischer-Tropsch (FT) catalysis on the surface
of solid iron grains. These FT reactions can convert CO and H$_2$
into water and hydrocarbons \citep{willacy04}, these hydrocarbons
could then be converted in into PAHs. 

So far it remains unclear whether the shape of the observed emission features can detect if
the PAH carriers reside in the disc or an outflow.

\section{Spectral energy distribution}
\label{sed}

\subsection{2D disc modelling}

For both stars extensive photometric data are available. This, together with the Spitzer infrared
spectral information, allows us to constrain some of the physical characteristics of the circumbinary disc.

We fit the SED using a Monte Carlo code, assuming 
2D-radiative-transfer in a passive disc model \citep{dullemond01,dullemond04}.
This code computes the temperature structure and density of the disc. The
vertical scale height of the disc is computed by an iteration process,
demanding vertical hydrostatic equilibrium. The distribution of dust grain properties
is fully homogeneous and, although this model can reproduce the SED,
dust settling timescales indicate that settling of large grains to the midplane occurs.
Using the dust settling timescale 
$$t_{set}=\frac{\pi}{2}\frac{\Sigma_0}{\rho_d a}\frac{1}{\Omega_k}\ln\frac{z}{z_0}$$ with $\Sigma_0$ the surface density, $\rho_d$ the particle
density, $a$ the grain size and $\Omega_k=\sqrt{\frac{GM_*}{r^3}}$ the
Keplerian rotation rate \citep{miyake95}, we find that grains with sizes
between 500\,$\mu$m and 0.1\,cm can descend 50\,AU on timescales
similar to the estimated lifetime of the disc. An inhomogeneous
disc model with a vertical gradient in grain-size distribution is thus
necessary \citep{gielen07}. These large and cooler grains in the disc
midplane mainly contribute to the far-IR part of the SED and constitute
the main fraction of the total dust mass. The disc
structure and near- and mid-IR flux are almost fully determined by
small grains. So we use a homogeneous 2D disc model to fit the near-
and mid-IR part of the SED and add a single blackbody temperature to
represent the cooler midplane made up of large grains. The lack of
observational constraints on the temperature structure of this
component does not allow us to constrain this extra parameter.

Stellar input parameters of the model are the luminosity, the total
mass (we assume the total gravitational potential to be spherically symmetric with a
total mass of $M=1\,$M$_{\odot}$), and $T_{\rm eff}$. The luminosity
(and thus the distance) for these sources is not well constrained so
we use values between $L=1000-5000\,$L$_{\odot}$, typical values for
post-AGB sources. Input disc parameters are $R_{\rm in}$ and $R_{\rm out}$, the different dust components, the total disc mass and the
power law for the surface-density distribution. Since we are not
dealing with outflow sources a power law $\alpha >-2$ is used.
The gas-to-dust ratio is kept fixed at 100.

The modelling is still degenerate, especially in parameters like
the outer radius and the total disc mass which can be easily interchanged,
without strongly influencing the SED. We use a dust mixture of amorphous and crystalline
silicates in grain sizes ranging of $0.1-20$\,$\mu$m, with a power law
distribution of $-3.0$, for the
homogeneous disc. For HD\,52961 the 850\,$\mu$m submillimetre data
points to the presence of extremely large grains in the disc, but
these large grains are assumed not to influence the near- and mid-IR
part of the SED, and will only be important for the blackbody
component. We use a value of 300\,AU for $R_{\rm out}$. 
For the inner rim we use the radius at which the temperature of the inner rim
equals about 1500\,K. This is a typical value for the dust sublimation temperature of 
silicates, although values as low as 1200\,K are sometimes also used.  The total SED energetics are then calculated, given a specific
inclination angle of the system.

We do not aim to reproduce the observed narrow features in the Spitzer
spectrum, since these features originate from an optically thin upper
layer of small grains at the disc surface. They can only be fitted
well using an inhomogeneous disc model with grain settling. Instead,
we want to model the observed general energetics of the SED spectrum,
thus the observed continuum and amorphous dust features.

\subsection{Comparison with interferometric data}

The circumstellar environment of HD\,52961 has been resolved using the
VLTI/MIDI instrument, with angular sizes in the N-band between 35\,mas
and 55\,mas in a uniform disc approximation \citep{deroo06}. EP\,Lyr
is too faint for current interferometric capabilities.

To compare our disc model of HD\,52961 with the MIDI data we made model images
from which we extracted visibilities, using the same projected
baseline lengths (40\,m/46\,m) and angles ($45^{\circ}$/$46^{\circ}$)
as the observations. The only free parameters here are the
inclination of the disc and the orientation angle of the system on the
sky. A range of inclinations which still fit the observed SED was
tested, in steps of 15$^{\circ}$. The orientation angle is varied
1$^{\circ}$ at a time. The result of this comparison can be found in Sect.~\ref{resultshd52961}.

\subsection{Results}

\begin{figure}
\vspace{0cm}
\hspace{0cm}
\centering
\resizebox{9cm}{!}{ \includegraphics{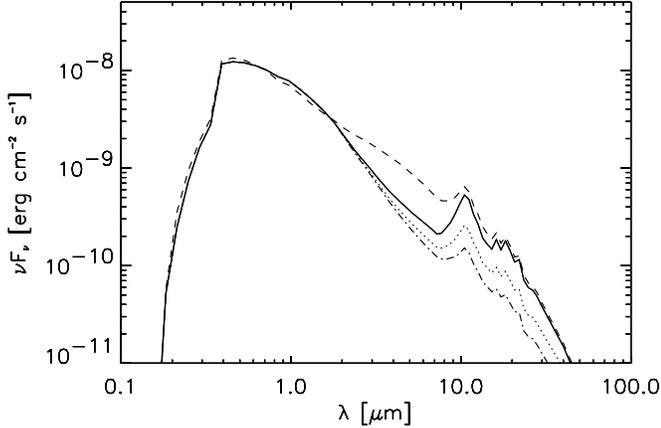}}
\caption{Some SED disc models showing the influence of the inclusion of metallic iron and different grain sizes
in the fitting. All models have the same physical parameters, the only difference being the amount of 
iron in the disc and the grain sizes. The solid line depicts a pure silicate disc with grain sizes between 0.1-20\,$\mu$m. 
The dashed line represents a model with 5\% metallic iron and 95\% silicate with grain sizes between 0.1-20\,$\mu$m.
The dotted line represents a silicate disc model with grain sizes between 0.1-50\,$\mu$m and the dot-dashed line
one with grain sizes between 0.1-100\,$\mu$m. }
\label{comparison_iron}%
\end{figure}

When modelling the near- and mid-IR part of the SEDs, the
feature-to-continuum ratio of the silicate features is 
too strong in comparison with the observed spectra for both stars. Moreover, the
near-IR flux is often underestimated. 
This was also
observed in \citet{gielen07}, where we fitted the SED similarly
to two post-AGB stars, RU\,Cen and AC\,Her. Including an extra
continuum opacity source is needed to reduce the strength of the
features (see Fig.~\ref{comparison_iron}) and to increase the near-IR
contribution. From our previous work \citep{gielen08} and Section~\ref{silicates}, we found
evidence that (a fraction of) the silicates might be iron poor.
So we use metallic iron as a potential
opacity source: its near-IR opacity is large, but the absorption
coefficient is unfortunately featureless so direct detection is
difficult. Inclusion of free metallic iron has a strong impact on the
modelling because the near-IR excess increases significantly with a
given inner radius. Another possible opacity source is the inclusion
of hot large grains in the homogeneous disc model.

\subsubsection{HD\,52961}
\label{resultshd52961}

\begin{table}
\caption{Results of our SED disc modelling. 
}
\label{sedresults}
\centering
\vspace{0.5cm}
\hspace{0.cm}
\begin{tabular}{lcccccc}
\hline \hline
          & $R_{\rm in} - R_{\rm out}$ & $m$         & $\alpha$ & iron & $T_{\rm bb}$  & $i$\\
          & AU                      & $10^{-5}$M$_{\odot}$ &          & \%   & K &   $^{\circ}$ \\
\hline 
HD\,52961: A   &  $10 - 500$  &  $1.7$ & -1.5 & 0  & 160  &  $0-90$\\
HD\,52961: B   &  $10 - 500$  &  $0.7$ & -1.5 & 10  & 160  &  $< 65$\\
EP\,Lyr: A  &  $40 - 300$  &  $6$   & -1.0  & 0  & -  &  $0-90$\\
\hline
\end{tabular}
\begin{footnotesize}
\begin{flushleft}
Note: Given are the inner and outer radius ($R_{\rm in}$-$R_{\rm out}$),
the total disc mass $m$ for the homogeneous disc model, the surface-density distribution power law $\alpha$, the percentage of iron 
in the homogeneous disc model,
the blackbody temperature and the inclination of the system.
\end{flushleft}
\end{footnotesize}
\end{table}  

\begin{figure}
\vspace{0cm}
\hspace{0cm}
\centering
\resizebox{8cm}{!}{ \includegraphics{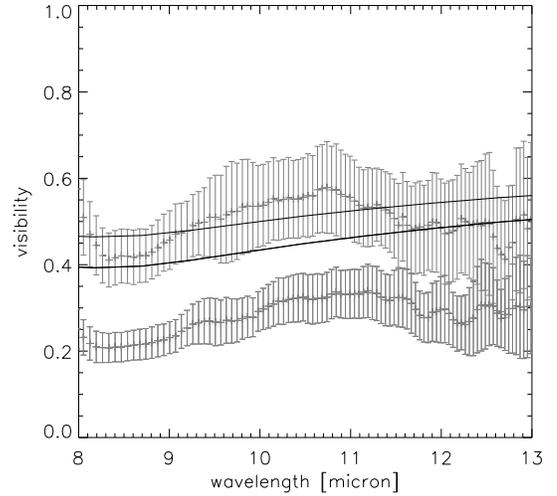}}
\caption{The observed MIDI measurements of HD\,52961 (gray data with error bars) and the visibilities deduced from disc model A (black solid line),
for the two different baselines.
The model has an inclination of 45$^{\circ}$ and a position angle on the sky of 227$^{\circ}$ East of North.}
\label{midimodel}%
\end{figure}

We can use the interferometric data for HD\,52961 to constrain the
distance to the system (Fig.~\ref{midimodel}). When we compare the
modelled visibilities with the observations for HD\,52961 we find that
the visibilities of our model are too high, when using standard
models of $L=5000\,$L$_{\odot}$ which fit the SED and impose the inner radius to be at sublimation radius. 
This means we need to increase the angular size of
the N-band emission region, either by increasing the physical size or
by decreasing the distance. Increasing the disc size to outer radii $>
500$\,AU does not influence the N-band emission so we need to increase
the inner radius, flatten the density distribution power law and/or decrease the luminosity in the disc model. 
Changing the surface-density distribution power law to values $>-1.5$ proved incompatible with the observed SED.
We therefore use an average luminosity of 3000\,L$_{\odot}$ and move
the inner radius to larger distances. A good fit to the SED was
obtained using an inner radius of 10\,AU. This assumed luminosity
gives a distance to the system of about 1700\,pc.
At 10\,AU the temperature of the inner rim is around 1100\,K, which is slightly below the
canonical dust sublimation temperature for silicates.

\begin{figure}
\vspace{0cm}
\hspace{0cm}
\centering
\resizebox{9cm}{!}{ \includegraphics{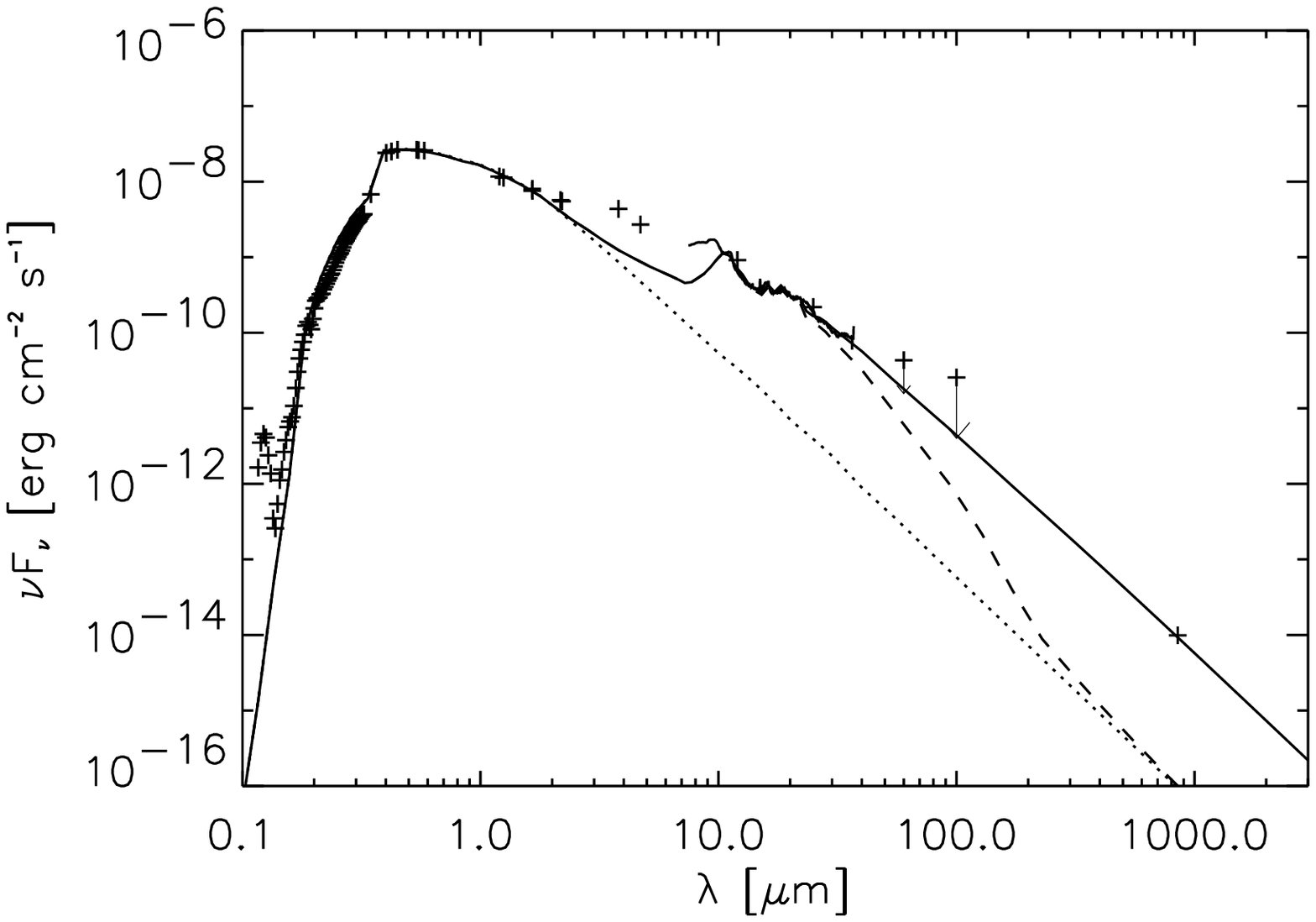}}
\resizebox{9cm}{!}{ \includegraphics{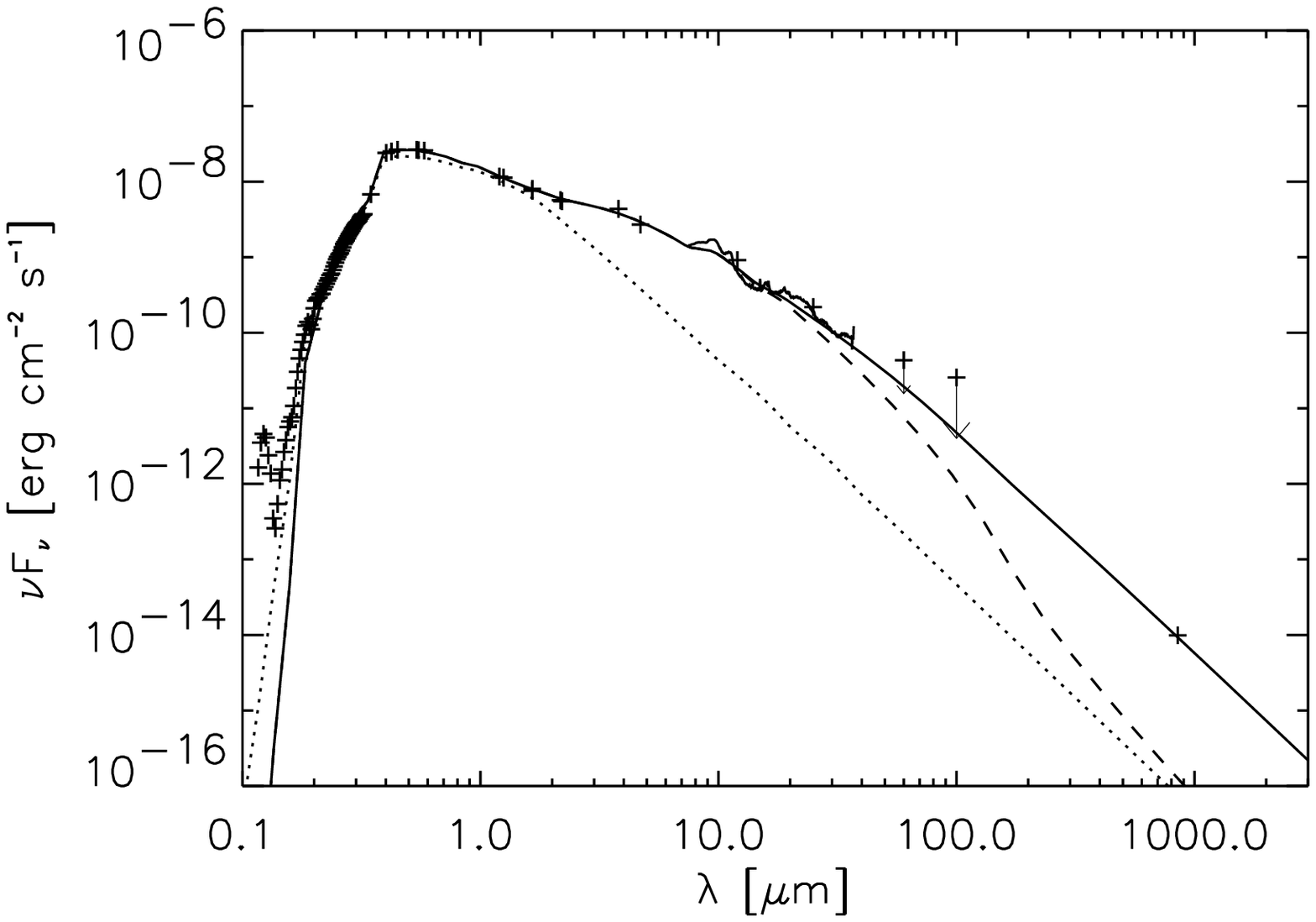}}
\caption{SED disc modelling of HD\,52961 (top: model A without metallic iron, bottom: model B with metallic iron). The dashed line represents the homogeneous disc model consisting of
grains between 0.1\,$\mu$m and 20\,$\mu$m. The solid line gives the disc model with an added blackbody
to represent the cool midplane. Crosses represent photometric data and in the infrared we overplot the observed Spitzer-IRS spectrum.
The dotted line represents the adopted Kurucz stellar model.}
\label{HD52961_sedfit}%
\end{figure}

For HD\,52961 we tested models with and without the inclusion of
metallic iron. The resulting fits can be seen in
Figure~\ref{HD52961_sedfit}, parameters can be found in
Table~\ref{sedresults}. Since this star shows a rather strong
10\,$\mu$m feature, pointing to relatively large amounts of hot dust
in the disc, we find we need a rather steep surface-density
distribution, $\alpha < -1$. Since we only have one flux point at long wavelengths, 
we add a simple 160\,K blackbody model to fit the
observed submillimetre emission. 

When no iron is included (model A) we see that the flux around
from 2 to 8\,$\mu$m is strongly underestimated. Including about 10$\%$ metallic
iron (model B) increased this flux significantly. The inclusion of
metallic iron has only a minor influence on the N-band interferometric measurements.
It decreases the modelled visibilities by about 10\%, which is still
consistent with the observed visibilities.

\begin{figure}
\vspace{0cm}
\hspace{0cm}
\centering
\resizebox{8cm}{!}{ \includegraphics{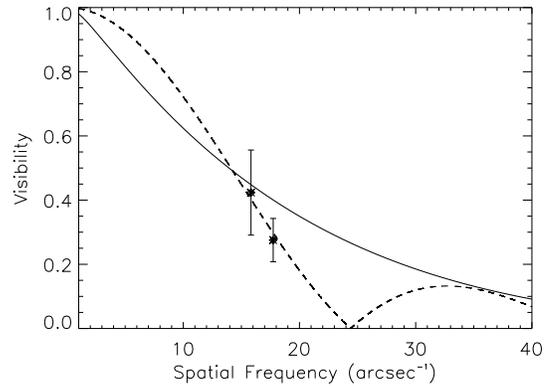}}
\caption{Comparison between modelled and observed visibilities at 12.6\,$\mu$m, at the two different baseline lengths. The dashed line represents a
uniform disc model with an angular size of 50\,mas. The solid line gives the visibilities as calculated
from the 2D disc modelling (model B).
}
\label{spatfreq}%
\end{figure}

The modelled visibilities (Fig.~\ref{midimodel}) lie within the observed visibility range but 
a remarkable detection is that the variation in visibilities between
the two observations is quite large, despite the very limited
difference in lengths ($41.3-46.3$\,m), as well as in projected angle
($45.6-46.3^\circ$). This is illustrated when we plot the visibility
versus the spatial frequency for a given wavelength, as seen in Figure~\ref{spatfreq}.
In this figure we illustrate that when using a uniform disc for
the intensity distribution, the steep visibility drop can be accounted
for. The physical disc model is, however, much smoother and does not
contain the very sharp edge characteristic of the uniform
disc. 
The model also do not reproduce the observed `bump' in visibility between 9 and 12\,$\mu$m.
\citet{deroo06} explain this observed increase in visibility as being due to a non-homogeneous distribution
of the silicates, which contribute most to the inner regions of the disc.
The current disc model does not include the physics to be able to
reproduce this radial distribution of dust species.

The submillimetre 850\,$\mu$m flux for HD\,52961 and the derived
blackbody temperature of 160\,K can be used to estimate the dust mass
of large grains in the disc. In the optically thin approach the disc
mass can be estimated using \citep{hildebrand83}
$$M_d=\frac{F_{850}\,D^2}{\kappa_{850}\,B_{850}(T)}.$$ 
Assuming a cross section of large spherical grains, the mass
absorption coefficient of 850\,$\mu$m grains in
blackbody approximation is about 2.4\,cm$^2$\,g$^{-1}$. The mass absorption coefficient
is given by $\kappa=\frac{\pi a^2}{\frac{4}{3}\pi a^3 \rho}$, with $a$ the grain size, and $\rho$ typically
3.3\,g\,cm$^{-3}$ for astronomical silicates. This results
in dust-mass estimates in large grains of $3.2\times
10^{-6}\,$M$_{\odot}$ for HD\,52961.

\subsubsection{EP\,Lyr}

The SED-fitting gives an estimate of the distance to the system of
$d=3.4$\,kpc, assuming a luminosity $L=3000\,$L$_{\odot}$. This
estimation is not only dependent on the assumed luminosity of
the star, but also on the adopted inclination of the system. Other
derived disc parameters can be found in Table~\ref{sedresults}.

For EP\,Lyr it proved very problematic to get a good fit to the
observed strong 20\,$\mu$m feature, without introducing a strong
10\,$\mu$m silicate feature. The mixed chemistry adds to the
complexity of the object as the PAH emission and underlying continuum
in the near-IR may very well come from a
distinct structural component, in for example the polar direction. 
This is already seen in HD\,44179, were the observed PAH emission comes from a
bipolar outflow \citep{bregman93,menshchikov02,cohen04}.
The lack of data shortwards of 7\,$\mu$m makes it hard to get a good continuum estimate of the near- and mid-IR
energetics. Cool dust clearly dominates the SED, but without additional information
from interferometry, we cannot constrain parameters like the inner radius or 
the surface-density distribution. We opted to keep a rather flat density distribution of $\alpha = -1.0$
and do not force the model to fit the spectrum shortward of 15\,$\mu$m.

To get a good fit to the SED a very large inner radius of about 40\,AU is necessary.
At this distance the inner rim reaches temperatures of $\sim300$\,K. This temperature agrees
with the temperatures found in the spectral modelling (Sect.~\ref{silicates}).
If we include metallic iron in the model we find we need inner radii even larger than 200\,AU,
which seems physically implausible. Unfortunately we do not posses submillimetre data for this
star so we cannot determine the blackbody temperature associated with the midplane.
The small $L_{\rm IR}/L_*$ as well as the  lack of a near-IR 
dust excess shows that the inner rim is quite far from the sublimation radius.

\begin{figure}
\vspace{0cm}
\hspace{0cm}
\centering
\resizebox{9cm}{!}{ \includegraphics{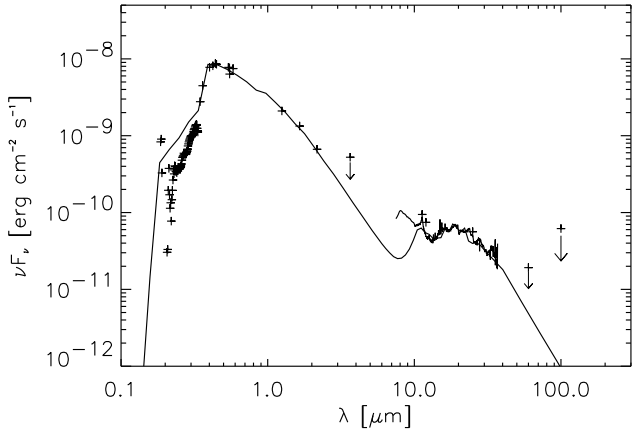}}
\caption{The SED disc model of EP\,Lyr. The model parameters are given in Table~\ref{sedresults}.}
\label{eplyr_sedfit}%
\end{figure}

\section{Conclusions} 
\label{conclusions}

HD\,52961 and EP\,Lyr both have rich infrared spectra, and the assembled
multi-wavelength data show that these evolved objects
are surrounded by a stable circumstellar disc. While the binary nature
of HD\,52961 is well established, the binarity of EP\,Lyr is suspected but not yet
firmly proven. The discs must be circumbinary
as the sublimation radius of the dust is larger than the determined (HD\,52961)
and suspected (EP\,Lyr) binary orbit.

Recent studies have shown that many of these binary post-AGB systems are already detected
\citep{vanwinckel03, deruyter06, deroo06, deroo07b,gielen08}, but EP\,Lyr
and HD\,52961 both show quite distinct characteristics in dust and gas
chemistry as well as in physical properties of their discs.

EP\,Lyr and HD\,52961 are the only stars from the larger Spitzer
sample that have clear CO$_2$ gas emission lines in
the mid-IR. Our modelling shows that the emission in both stars can be
well fitted and is dominated by $^{12}$C$^{16}$O$_2$ features,
but clear detections of other isotopes are present as well.
Similar excitation temperatures and column densities are found in both
objects, but with different ratios for $^{12}$C$^{16}$O$_2$ and
$^{13}$C$^{16}$O$_2$.
Why these two stars are the only ones from the larger sample showing
strong CO$_2$ features, and if there is any relation with the low
observed infrared flux remains unclear. Similar feature strengths observed
in the other stars would have been easily detected.
One possibility is that the low dust emission in the two sample stars, reflected in the low
$L_{\rm IR}/L_{star}$, makes it easier for the CO$_2$ gas to become visible.
This effect is also seen in AGB stars, where CO$_2$ gas emission is strongest
in sources with the lowest mass-loss rates \citep{sloan96,cami02,sloan03}.

One of the most remarkable features is the clear detection of $^{18}$O isotopes 
of CO$_2$ in both objects. Together with HR 4049 \citep{cami01}, the strong 
$^{16}$O$^{12}$C$^{18}$O band is a systematic feature of the gas emission in 
the discs of post-AGB binaries when CO$_2$ emission is detected. The high $^{18}$O 
abundance of HR\,4049 derived in an optically thin approximation of a putative 
nucleosynthetic origin \citep{lugaro05} was not confirmed by the analysis of 
CO first overtone absorption \citep{hinkle07} in the same object. It is likely 
that the CO$_2$ gas is strongly optically thick, also in EP\,Lyr and HD\,52961, 
so that very rare isotopes are detectable.

The high-resolution Spitzer spectra also reveal unique solid state
features. As in the bulk of the disc sources \citep{gielen08},
crystalline silicate features prevail in both stars, but unlike what we
found for the larger sample, they proved very hard to model.
In HD\,52961 we observe some unique strong crystalline
features at 11.3 and 16\,$\mu$m, which could not be reproduced in the
modelling, irrespective of the grain size used in the models, shape or
assumed grain model. In the 2D disc modelling we could not fit the steep rise
around 10\,$\mu$m, without the inclusion of metallic iron. Combining
our physical model constrained by the SED, together with our
interferometric data, we concluded that the inner dust rim is slightly beyond
the dust sublimation radius. This is in contrast to similar binary
objects like IRAS\,08544-4431 where the interferometric data shows that
the dusty disc has to start very near to the dust sublimation radius
\citep{deroo07b}. Assuming a luminosity of 3000\,L$_{\odot}$,
we find that the inner disc radius of HD\,52961 is rather large,
around 10\,AU. The strong 850\,$\mu$m flux shows that this object has a
component of very large grains. This contribution was added to the SED fitting by
an additional colder Planck curve.

EP\,Lyr has only a very small infrared excess, but the Spitzer
spectrum is very rich in spectral details. The most remarkable
characteristic is the clear PAH emission, in combination with the
strong crystalline features at longer wavelengths. There is no
evidence that the central star evolved into a carbon star when on the
AGB, yet unprocessed class-C PAH features are clearly detected.
Whether these PAH species are formed in the circumbinary disc or in a
recent, likely polar outflow of the depleted central star, remains
unclear. An extra component of cold dust is
necessary in this object as well, to fit the entire Spitzer spectrum. Unfortunately,
EP\,Lyr is too faint for the current interferometric possibilities.

The mixed chemistry, the strongly processed cold crystalline silicates and low 
$^{12}$C/$^{13}$C ratio are in common with the subgroup of silicate J-type carbon
stars, which can also display strong crystalline material.
This corroborates the conclusion that in the latter, the disc is circumbinary.
The abundance studies of J-type carbon stars are not complete enough to
probe whether photospheric depletion affected these objects as well.

Both objects are extreme examples of post-AGB binary stars, with
characteristics dominated by the presence of a stable
circumbinary disc. This disc environment is, to first order, well
modelled by assuming a passive, irradiated stable disc.
In this paper we corroborate that this geometry is ideal to induce
strong grain processing and a rich, even mixed chemistry.
We conclude also that a homogeneous disc model is too primitive to
model the spectral details as evidence of grain settling is strong.
The route to PAH formation (and excitation) in the O-rich EP\,Lyr
remains to be studied in detail as PAH emission is only observed in a
very limited number of such sources.

This detailed study of HD\,52961 and EP\,Lyr shows that many questions
still remain in our current understanding of the evolution of a significant number of post-AGB binary stars,
and the impact of the circumbinary discs on the entire system.

\begin{acknowledgements}
CG acknowledges support of the Fund for Scientific Research of Flanders
(FWO) under the grant G.0178.02. and G.0470.07.
\end{acknowledgements}

\bibliographystyle{aa}
\bibliography{/STER/100/cliog/disk28/Artikels/referenties}

\end{document}